\documentstyle[aas2pp4]{article}

\begin{document}
\slugcomment{accepted to the Astrophysical Journal}

\title{Galaxy Selection and Clustering and Ly$\alpha$ Absorber Identification}
\author{Suzanne M. Linder$^1$}
\affil{The Pennsylvania State University, University Park, PA 16802}
\authoremail{slinder@inaoep.mx}
\altaffiltext{1}{Current address:  INAOE, Apartado Postal 51 y 216,
Puebla 72000, Pue., Mexico}

\begin{abstract}
The effects of galaxy selection on our ability to constrain the nature
of weak Ly$\alpha$ absorbers at low redshift are explored.  
Current observations indicate the existence of
a substantial population of gas-rich, low
surface brightness (LSB) galaxies, and these galaxies may have large 
cross sections for Ly$\alpha$ absorption.
Absorption arising in LSB galaxies is likely to be attributed to high surface 
brightness galaxies at larger impact parameters from quasar lines of 
sight, so that the observed absorption cross sections of galaxies may seem 
unreasonably large.  Thus it is not currently possible to rule out scenarios
where LSB galaxies make substantial contributions to Ly$\alpha$ absorption
using direct observations.
Less direct tests, where observational selection effects are taken into
account using simulations, should make it possible to determine the nature
of Ly$\alpha$ absorbers by observing a sample of $\sim 100$ galaxies around
quasar lines of sight with well-defined selection criteria.
Such tests, which involve comparing simulated and 
observed plots of the unidentified absorber fractions and absorbing galaxy
fractions versus impact parameter, can distinguish between
scenarios where absorbers arise in particular galaxies and those where
absorbers arise in gas that traces the large scale galaxy distribution.
Care must be taken to minimize observational selection effects even when
using these tests.  Results from
such tests are likely to be dependent upon the limiting absorption line 
equivalent width or neutral hydrogen column density.  While not enough data
are currently available to make a strong conclusion about the nature of 
moderately weak absorbers, some evidence is seen that such absorbers arise
in gas that is around or between galaxies that are often not detected in
surveys.
\end{abstract}


\section{Introduction}
Ly$\alpha$ forest absorbers, seen shortward of Ly$\alpha$ emission in quasar
spectra, are powerful tools for studying the formation and evolution
of galaxies and large scale structure.  With the ultraviolet capabilities of
the Hubble Space Telescope it has become possible to observe these absorbers
at redshifts spanning the entire range from $0<z<5$, including those
at low enough redshifts (Bahcall et al.~1996; Weymann et al.~1998; Jannuzi et
al.~1998) so that many possibly associated galaxies can be
found close to the quasar lines of sight.  Using absorbers to study galaxy
evolution will require knowing what fraction of absorbers arise in galaxies
and what kinds of galaxies give rise to absorption.  Yet it has been 
difficult to 
establish the nature of these weak, low redshift absorbers using direct
observations because the absorbers do not generally arise near the luminous 
parts of galaxies that are typically seen in surveys.

The question thus remains as to whether weak Ly$\alpha$ absorbers are 
associated with particular
galaxies or whether they arise in gas that traces the large scale galaxy
distribution.  The major emphasis here will be placed on finding the nature
of absorbers between $10^{17.2}$ cm$^{-2}$ and $10^{14.3}$ cm$^{-2}$, although
the same methods will be useful in constraining the nature of weaker absorbers.
The definition of absorbers being `associated' with galaxies
is not clear.  Some absorbers could arise in gas which is gravitationally
bound to galaxies, but here `associated' simply means that, on average, 
there is some falloff of neutral hydrogen column density with distance from
the centers of galaxies.  It has been shown (Linder 1998) that galaxies could
have sufficient absorption cross sections to explain all of the Ly$\alpha$
absorbers, assuming that low surface brightness (LSB) galaxies (see Bothun,
Impey, \& McGaugh 1997) are included.  Other studies have suggested that 
absorbers arise largely in luminous, high surface brightness (HSB) galaxies
(Chen et al.~1998; Lanzetta et al.~1995).  However, most recent studies
(Bowen, Blades, \& Pettini 1996; Bowen, Pettini, \& Boyle 1998; 
Dav\'e et al.~1999; Le Brun, Bergeron, \&
Boiss\'e 1996; Morris et al.~1993;
Shull, Stocke, \&
Penton 1996; Tripp, Lu, \& Savage 1998; van Gorkom et al.~1996) have argued
that absorbers trace the large scale distribution of galaxies, rather than
being associated with the luminous HSB galaxies which are seen in the
surveys.  The most 
common test used in attempt to establish the nature of absorbers has
been looking for an anticorrelation between the absorption equivalent width
and the impact parameter between a galaxy and quasar line of sight.  However,
the presence (or lack) of such an anticorrelation has been used to support
a variety of viewpoints.  While absorbers are generally found to trace
the large scale galaxy distribution, some weak 
absorbers have also been found in cosmic voids
(Shull et al.~1996).  It is interesting to note here that LSB galaxies may be
more weakly clustered than HSB galaxies (see below), so that they are more 
likely to be able to give rise to absorption in void regions.

Whether absorbers arise largely in LSB galaxies or in gas between the galaxies,
it would be difficult to design a direct observational test to distinguish 
between these possibilities.  An observer is only capable of finding galaxies
close to, or within some angular separation or impact parameter of, a quasar
line of sight which satisfy some selection criteria.  Given the possibility
that LSB galaxies could make a substantial 
contribution to absorption, understanding and simulating observational 
selection effects will be crucial in testing any model for the nature of
Ly$\alpha$ absorbers.  For example, absorption arising in unseen LSB galaxies
may be attributed to luminous HSB galaxies at larger impact parameters from
the quasar line of sight,  so that the absorption cross sections of the HSB 
galaxies may be overestimated.  

Since observational tests for the nature of weak Ly$\alpha$ absorbers are
likely to involve measuring impact parameters between galaxies and quasar
lines of sight, the results of such tests depend upon the clustering
properties of galaxies and of gas around galaxies.  There is some evidence 
that LSB galaxies are more weakly clustered than HSB galaxies.
While they have not been found to fill the voids, they appear to have a
lack of close companions (Bothun et al.~1993) and a lower amplitude in the 
LSB-HSB correlation
function compared to that for the HSB autocorrelation function (Mo, McGaugh,
\& Bothun 1994).  While
most absorbers are found in the same regions where galaxies are located,
evidence is also seen that some rich galaxy clusters do not give rise to
absorption (Tripp et al.~1998).

Recently cosmological simulations have been used to investigate 
low redshift Ly$\alpha$ absorbers (Dav\'e et al.~1999).  While
such simulations allow for a more sophisticated treatment of some 
physical properties of absorbers, the simulations
used here, which are described further in Linder (1998), have the advantage
of being more clearly connected to the observable properties of galaxies.
The lack of LSB galaxies produced by cosmological simulations such as Dav\'e
et al.~make such simulations less conclusive in determining the relationship
between galaxies and absorbers.
Furthermore, simulating observational selection effects will be necessary
in testing the results of any simulations.

Section 2 describes the simulation methods used, and Section 3 describes
the effects of galaxy selection upon tests currently used to find the
nature of Ly$\alpha$ absorbers.  In Section 4, I propose some possibly 
more conclusive tests and discuss the use of these tests for
distinguishing between various scenarios for low redshift Ly$\alpha$ 
absorbers, including galactic and nongalactic absorbers.  In Section 5
the proposed tests are applied to some currently available observations,
and complications in using such tests are discussed. It is assumed
that the value of $H_0=100$ km s$^{-1}$ Mpc$^{-1}$. 

\section{Method}

Simulated observations with realistic selection criteria are made of 
simulated galaxies, which give rise to absorption, in order to 
investigate effects of galaxy
selection.  In the standard scenario, all absorbers with $N_{HI}>10^{14.3}$ cm$^{-2}$
arise in extended galaxy disks, and most are associated with LSB galaxies.
Simulated observations are made of various other scenarios,
described in Section 4, to investigate the possibility of 
observationally distinguishing between such scenarios.  Since an observer
cannot conclusively identify which galaxy (if any) gives rise to weak 
absorption, the closest galaxy to a quasar line of sight which
satisfies some selection criteria is identified for each absorber.  For
absorbing galaxy fraction (AGF, or the fraction of galaxies that appear to
give rise to absorption within some impact parameter) plots, an observer is
likely to assume that multiple galaxies contribute to an absorption line
which could contain multiple unresolved components.  Thus any galaxy is
considered to give rise to absorption if an absorption line is found within
some velocity difference (400 or 750 km/s, depending on the data set to
be compared).

\subsection{The Standard Scenario}\label{standard}

The `standard scenario' refers here to one where absorbers arise largely 
around LSB galaxies, as was simulated in Linder (1998). 
The simulation used for the standard scenario
is described as Case 10 in Linder (1998), where weak absorbers arise in
ionized gas that extends from galaxy disks, as modeled in Charlton, Salpeter,
\& Hogan (1993) and Charlton, Salpeter \& Linder (1994).  Each galaxy has
an ionized outer disk in which the neutral column density declines as a
power law with radius.
The number of simulated galaxies
and box size are chosen in a manner that facilitates using the clustering 
simulation described below and so that a realistic value is produced for
low redshift absorber counts.  The average number of absorbers per unit
redshift along a line of sight at redshift zero
was found by Bahcall et al.~(1996) to be
 $(dN/dz)_0=24.3\pm 6.6$, complete to an equivalent width in
Ly$\alpha$ of 0.24 \AA.

Instead of placing galaxies randomly within the box, the galaxies are given
clustered positions that are chosen using the method of Soniera \& Peebles
(1978) with between seven and nine clustering levels (eight for the standard
scenario).  In this method, richer clusters have more clustering levels,
such that a cluster with $l$ levels contains $2^l$ galaxies.
In order to make LSB galaxies more weakly clustered, the galaxies
are moved outward from the centers of the second clustering level by a factor
of $1+0.425(\mu_B(0)-21.65$ mag arcsec$^{-2})$, for a galaxy with central 
surface brightness $\mu_B(0)$. This factor is chosen so that LSB galaxies
are located, on average, about twice as far from the the cluster centers as
HSB galaxies, in order to reduce the amplitude of the LSB-HSB correlation 
function.  For the standard scenario, 16384 galaxies are placed in a
cube with an edge of 28.6 $h^{-1}$ Mpc.  The amplitude of the autocorrelation function
for HSB galaxies is 5.6 $h^{-1}$ Mpc, which is 2.3 times larger
than that for the LSB-HSB autocorrelation function, where LSB galaxies are
defined to have central surface brightnesses 
$>23.06$ $B$ mag arcsec$^{-2}$.

\subsection{Alternate Scenarios}

To examine the sensitivity of the results to the assumptions in the standard
scenario, several models were created.
Different scenarios include a galactic halo absorber model and scenarios
where absorption arises in gas that is clustered around luminous HSB 
galaxies.  Variations are also made in the clustering behavior of galaxies,
the shape of the galaxy central surface brightness distribution, the
ionizing background radiation, and the
slope, $t$,  of the `Holmberg' relation between galaxy absorption radius and 
luminosity, where $R/R^*=(L/L^*)^t$ and $R^*$ is the absorption radius at a
neutral column density cutoff of $10^{14.3}$ cm$^{-2}$ for an $L^*$ galaxy.  
The effects of making the gas clumpy are also explored.

As an alternative galactic absorber model, the standard scenario is compared
with galactic absorbers obeying equation (24) of Chen et al.~(1998).
The same luminosity and surface brightness distributions and clustering 
behavior as for the standard scenario are used for the galaxies here.
The major differences between this scenario and the standard scenario are
that absorption arises in galaxy halos rather than extended disks and 
that the absorption cross sections of galaxies are unrelated to surface
brightness.   There is also a small difference in the slope of the Holmberg
relation:  $t=0.4$ compared to $0.5$ in the standard scenario.  
The number density of galaxies is changed by a factor of
$0.65$ in order to produce the observed absorber counts. The difference is
largely due to
the spherical rather than disk absorber geometry.  Note that in this 
scenario, about half of the absorbers still arise in LSB galaxies.

Numerous observational studies have reported that absorbers arise in gas
that traces the large scale galaxy distribution.  In other words, weak
absorption is seen to arise within several hundred kpc of the luminous HSB
galaxies that are detected in such surveys.  Thus two scenarios are 
simulated in which stronger ($N_{HI}>10^{16}$ cm$^{-2}$) absorbers are
associated with galaxies, while weaker absorbers arise whenever a line
of sight passes within 500 kpc (or 750 kpc) of a galaxy with $M_{B}<-19$
($-18$) and $\mu_B(0)<22$ mag arcsec$^{-2}$.  The cluster sizes are 
chosen to produce observed absorber counts, and the neutral column densities
for the nongalactic absorbers are chosen from a power law distribution:
$f(N_{HI})\propto N_{HI}^{-1.5}$, where $N_{HI}$ 
is between $10^{14.3}$ cm$^{-2}$ and $10^{16}$ cm$^{-2}$.
The absorbers are not clustered around LSB ($\mu_B(0)>22$ mag arcsec$^{-2}$)
galaxies here because absorbers are generally reported to be clustered
around easily visible galaxies, while LSB galaxies do not trace the 
large scale galaxy distribution as strongly.  It is possible that 
absorption also arises in gas surrounding LSB and/or dwarf galaxies, but
then the scenario would have more resemblance to a galactic absorber scenario,
even if the gas is clumpy as discussed below.
Another possibility is that some absorbers arise in discrete, possibly 
large, clouds which are clustered in a similar manner as galaxies.  For
example, some high velocity clouds are thought to be located at large within the
local group (Blitz et al.~1999), although little is known about the general
properties of extragalactic high velocity clouds.  At this time there is
no evidence that such a scenario would be distinguishable from one in 
which extreme LSB galaxies give rise to substantial absorption.

Recently evidence has been reported that the galaxy central surface brightness
distribution is lognormal at a given luminosity (de Jong \& Lacey 1999).
While it is unclear how these $I$-band observations relate to what would be
seen in $B$ (which is simulated here) without extinction, a surface brightness
distribution which is lognormal in $B$ can be simulated.  Compared to the
flat distribution (McGaugh 1996) used in the standard scenario, this surface
brightness distribution would allow for fewer extremely large, Malin-type
LSB galaxies.  However, the majority of absorbers arise in galaxies that
are moderate in luminosity and surface brightness in Linder (1998) and the
standard scenario.  Thus most absorbers still arise in LSB galaxies when the
flat surface brightness distribution (at a given scale length) is replaced
by one which is lognormal at a given luminosity.  Such a 
distribution is simulated assuming
\begin{equation}
\Phi(\mu_0,M_B) = \frac{1}{\sqrt{2\pi\sigma_\mu}}\exp -\frac{(\mu_0-\mu^*)-
(M_B-M_B^*)
/3}{2\sigma_\mu^2},
\end{equation}
with $\sigma_\mu=0.65$ and $\mu^*=22.65$ $B$ mag arcsec$^{-2}$, 
where a Schechter 
luminosity function is obeyed as in the standard scenario.
A simulation is also made with $\mu^*=21.65$ $B$ mag arcsec$^{-2}$, as the 
value is quite uncertain.  With this $\mu^*$ value, an extreme scenario is
produced where the majority of galaxies are compact and high in surface 
brightness.  Furthermore, an excessive number of Lyman limit systems are
produced, as in some cases discussed in Linder (1998).

Some attempts are made to vary the clustering behavior of galaxies, although
only some combination of discrete numbers of clustering levels can be used
for the Soniera \& Peebles (1978) method.  In addition to the standard
scenario with eight levels of clustering, two others are made with 
half of the
galaxies at level 7 (or 9) and the remaining galaxies at level 8.  In both cases
changes in the correlation functions 
occur largely due to the small box size used in the simulations here.
While Soniera \& Peebles included some richer clusters in their simulated
galaxy distribution, these make up a small fraction of the clusters, so
that these rich clusters would occur very infrequently given the small box
size and number density of galaxies typically simulated here.  Furthermore,
some evidence is seen (Tripp et al.~1998) that the richest clusters may
be less likely to give rise to absorption, possibly due to increased 
amounts of ionizing radiation in such clusters.

The gas surrounding galaxies is made clumpy by allowing
the neutral column density expected in the standard scenario
to vary by up to two orders of magnitude.  
Little change is seen, compared to the standard scenario, in the test
results described below, as it is only required that the neutral column 
density falls off, on average, with distance from the centers of galaxies.
It is also possible that low luminosity galaxies make a larger contribution
to absorption than in the standard scenario.  The slope of the Holmberg
relation can be varied in order to increase absorption cross sections
for dwarfs relative to luminous galaxies.  A simulation is made with 
absorption radius $R\propto L^{0.4}$ as reported by Chen et al.~(1998).
Finally, as an extreme case, a random weak absorber scenario is simulated
by choosing random new positions for absorbers $<10^{16}$ cm$^{-2}$ produced
in the standard scenario while the stronger absorbers remain associated with
galaxies.

In summary, a list of the simulated scenarios is given here:
\begin{enumerate}
\item The standard scenario is described in Sec.~\ref{standard}.  In this
scenario, most absorbers arise in gas extending from disks of somewhat
luminous LSB galaxies.
\item The Chen et al. (1998) scenario has observable properties of galaxies
that are the same as in the standard scenario, but the absorption properties
are described by equation (24) of Chen et al.
\item A nongalactic scenario is simulated where absorbers $<10^{16}$ cm$^{-2}$
arise when a line of sight passes within 750 kpc of an HSB ($\mu_B(0)<22$ mag
arcsec$^{-2}$) galaxy with $M_B<-19$.  Observable properties of galaxies 
remain the same, and stronger absorbers arise in galaxies, as in the standard
scenario.
\item A second nongalactic scenario is simulated which is identical to the
one above (3) except that absorbers $<10^{16}$ cm$^{-2}$ arise within 500 kpc
of an HSB galaxy with $M_B<-18$.
\item The clustering behavior of galaxies was varied, so that the observable
and absorption properties of galaxies remained the same as the standard
scenario, but half of the galaxies were moved from clustering level 8 to level
7.
\item The clustering behavior was varied as in (5) above, but half of the
galaxies were moved to level 9.
\item A lognormal surface brightness distribution was simulated where $\mu_B^*=
22.65$ mag arcsec$^{-2}$.  Absorption and other observable galaxy properties 
remain as in the standard scenario.
\item The Holmberg relationship was changed to $R\sim L^{0.4}$ from $R\sim L^
{0.5}$ for galaxy absorption cross sections.  Observable properties of galaxies
remain the same as those in the standard scenario.
\item  Clumpy gas is simulated by allowing column densities found in the 
standard scenario to vary randomly by up to two orders of magnitude.
\item  Random weak absorbers are produced by assigning new, random positions to
absorbers $<10^{16}$ cm$^{-2}$.  Stronger absorbers remain associated with
absorbers as in the standard scenario.
\end{enumerate}

\section{Effects of Galaxy Selection}

\begin{figure}[tb]
\plotfiddle{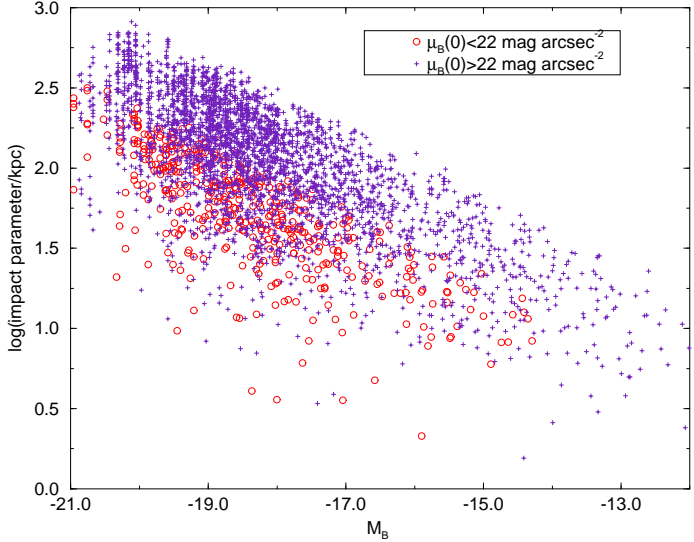}{2.3in}{270}{45}{45}{-185}{255}
\vskip -0.1in
\figcaption[fig1.eps]{The impact parameter (between the galaxy center and 
the quasar
line of sight) is plotted versus absolute magnitude $M_B$ 
for the actual absorbing ($>
10^{14.3}$
cm$^{-2}$) galaxies simulated
in the standard scenario.  More luminous galaxies have larger absorption radii,
and can thus cause absorption at larger impact parameters from the line of
sight.  LSB galaxies have larger absorption radii than HSB galaxies at a
given luminosity.  Vertical lines appear because luminous galaxies are able
to give rise to absorption along multiple simulated lines of sight.
\label{fig1}}
\end{figure}

A wide range of limiting luminosities have been used in surveys for galaxies
around quasar lines of sight.  Such limiting luminosities are likely to vary 
with distance to the galaxies and thus with the redshift range in which 
absorption lines can be detected.  Galaxies are typically identified if they
have some minimum diameter within some limiting isophote (McGaugh, Bothun,
\& Schombert 1995; Disney \& Phillips 1983), so that selection
biases are strongest against LSB galaxies and compact dwarf galaxies.  LSB
galaxies are often quite large in optical size, so that it is reasonable
to expect that they would also have large absorption cross sections, as in
Linder (1998).  Relatively few surveys have looked for LSB galaxies around
quasar lines of sight (van Gorkom et al.~1996; Chen et al.~1998).  The 
absorption cross sections of dwarf galaxies are likely to vary considerably.  
Some evidence is seen that galaxy absorption cross sections increase with
luminosity (Chen et al.~1998) although some especially extended dwarfs may
exist (van Gorkom et al.~1996).  For the standard scenario, it is
assumed that absorption cross sections increase with galaxy luminosity,
where $R\propto L^{0.5}$, which is similar to the relationship reported by 
Chen et al.  Since LSB galaxies are assumed to have larger absorption 
cross sections at a given luminosity, they are able to give 
rise to absorption at larger impact parameters from quasar lines of sight,
as seen in Fig.~\ref{fig1}.

\begin{figure}[tb]
\plotfiddle{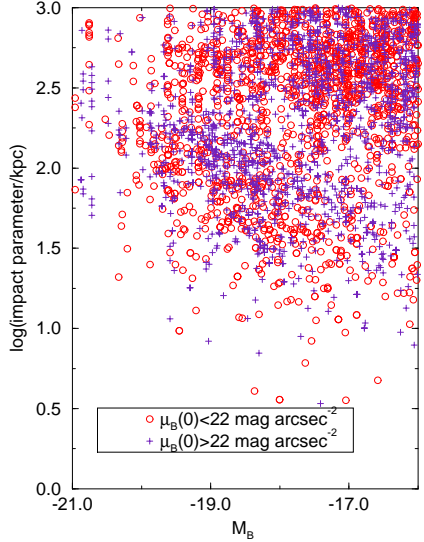}{2.3in}{270}{45}{45}{-185}{255}
\vskip -0.1in
\figcaption[fig2.eps]{The impact parameter versus $M_B$ are shown as if they were obtained
by `observing' the simulated galaxies from the standard scenario, in order to 
identify an absorbing
galaxy, according to the selection criteria described in Sec~ 3 ($M_B<-16$,
$\mu_B(0)<23$ mag arcsec$^{-2}$, $\triangle V<400$ km s$^{-1}$). While the
same simulated galaxies shown in Fig.~1 are `observed' here, it
is no longer obvious that LSB galaxies have larger absorption cross sections.
\label{fig2}}
\end{figure}

Suppose that the simulated sample of galaxies from Fig.~\ref{fig1} is 
`observed' with
the following selection criteria:  For each absorption line, the nearest
galaxy to a line of sight is found within a velocity difference of 400 km
s$^{-1}$, where the galaxy has $M_B<-16$ and $\mu_ B(0)<23$ mag arcsec$^{-2}$.
Again $M_B$ is plotted versus the impact parameter for the 'observed' 
absorbing galaxy in Fig.~\ref{fig2}. 
It is likely that an observer could find a way 
of determining that the points in the upper right-hand corner for 
Fig.~\ref{fig2}
are unphysical absorber-galaxy associations (Lanzetta, Webb, \& Barcons 1998).
For the remaining points, a correlation
between luminosity and impact parameter can still be seen.
Absorption arising in LSB galaxies is frequently `observed' as
arising in HSB galaxies at typically larger impact parameters from the
quasar line of sight.  Thus it is no longer possible to verify from the
`observed' plot that LSB galaxies have larger absorption cross sections
than HSB galaxies,
as assumed in the simulation.  With such reasonable selection criteria,
it may be possible for an observer to detect a relationship between 
galaxy absorption cross section and luminosity (although the slope may
be difficult to measure), but it is not  easily possible to 
detect a relationship
between absorption cross section and surface brightness.  For a sample
of 200 galaxies, a difference in the distributions for LSB ($>22$ $B$ mag
arcsec$^{-2}$) and HSB impact parameters, can
be detected with a K-S test about 50 percent of the time.  Thus while
Chen et al.~(1998) report no evidence for a relationship between galaxy
surface brightness and absorption cross section, it would not be possible
to detect such a relationship at this time.

Note that in actual surveys, the selection criteria are not always 
well-defined, although it should be possible to simulate any selection 
criteria that are clearly defined.  While surveys such as Chen et al.~(1998)
and Bowen et al.~(1996) include some galaxies as faint as $M_B=-16$,
they are not complete to this limit.  Observing more galaxies that are 
fainter or lower in surface brightness may make it easier to detect a 
difference in the LSB and HSB impact parameter distributions, but this 
may be difficult if there are not enough absorbers that are sufficiently
low in redshift to observe.

\begin{figure}[tb]
\plotfiddle{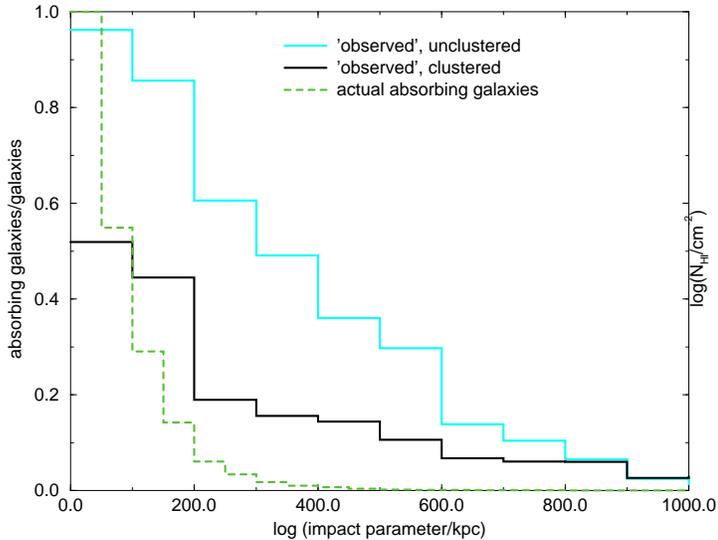}{2.3in}{270}{45}{45}{-185}{255}
\vskip -0.1in
\figcaption[fig3.eps]{A `covering factor' plot shows the fraction of galaxies found
at some impact parameter from a line of sight that apparently give rise to
absorption
($>10^{14.3}$ cm$^{-2}$).  Standard scenario galaxies are `observed' with
the same selection criteria as in Fig.~2.  Galaxy absorption radii are 
likely to be overestimated
when the galaxies are observed with strong selection effects against LSB
galaxies.  Galaxy clustering also causes misidentification of the actual
absorbing galaxy to occur more frequently.
\label{fig3}}
\end{figure}

A plot frequently made by observers (for example, Bowen et al.~1996) is that 
illustrating
the absorption covering factor, or the fraction of galaxies found to
cause absorption as a function of impact parameter, as shown in 
Fig.~\ref{fig3}.
`Observed' galaxies appear to cause absorption at large impact parameters
compared to those seen in Fig.~\ref{fig1}.  Again it can be seen that 
many absorbers
arising in LSB galaxies are attributed to HSB galaxies at larger impact
parameters from the quasar line of sight.   It can also be seen that
clustered absorbing galaxies are more frequently misidentified, as an
observed must choose from more galaxies that are close to an absorption
line.  Note that observers often assume that multiple galaxies can 
contribute to an absorption line, as is done in the following section.
This causes the covering factor to be even larger.

\begin{figure}[tb]
\plotfiddle{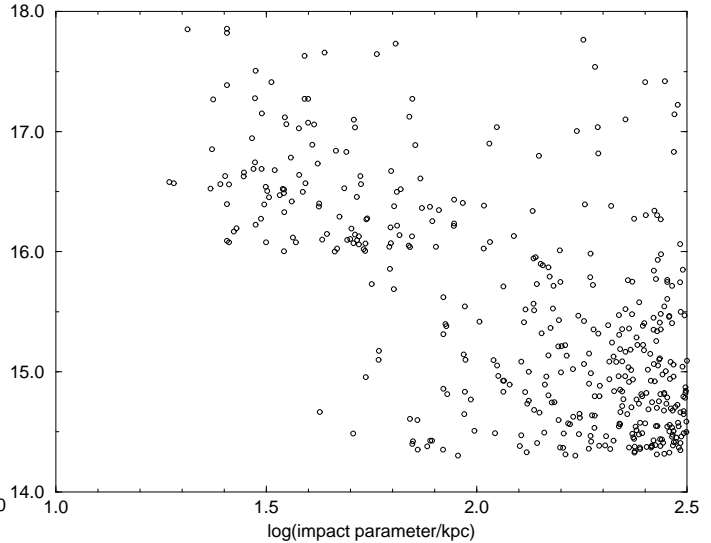}{2.3in}{270}{45}{45}{-185}{255}
\vskip -0.1in
\figcaption[fig4.eps]{Neutral column densities are plotted versus 
impact parameter
for simulated, absorbing ($>10^{16}$ cm$^{-2}$) galaxies (as in the standard
scenario) and
randomly distributed weaker absorbers (scenario 10), where the galaxies are `observed' as
above (but for $M_B>-18$).
An anticorrelation can arise even if weaker absorbers are not associated
with galaxies in any way.
\label{fig4}}
\end{figure}

The test most commonly used to establish the nature of Ly$\alpha$ absorbers
has been looking for an anticorrelation between equivalent width or 
neutral column density and impact parameter between the galaxies and quasar
lines of sight.  In Fig.~\ref{fig4}
it is shown that such an anticorrelation will 
arise even if absorbers $<10^{16}$ cm$^{-2}$ are distributed randomly 
relative to galaxies and not associated with galaxies in any way.  While
it is well established that some stronger Ly$\alpha$ absorbers are 
correlated with galaxies, the nearest observable galaxy to a line of sight 
is likely to be located at an impact parameter of a few hundred kpc whether
it gives rise to absorption or not.  Mostly unseen dwarf galaxies give rise to 
absorption at smaller impact parameters, whereas surveys typically look
out to a several hundred kpc when looking for luminous HSB galaxies.
Therefore the presence of such an
anticorrelation tells us little about the nature of weak Ly$\alpha$ absorbers.

\section{Tests for Distinguishing Between Scenarios}

Finding a relationship between absorption properties and observable 
galaxy properties would support the idea that absorbers are associated
with galaxies.  However, establishing such a relationship may be difficult
using such direct tests, as was shown in the previous section.  Here, other
tests that may be more efficient are explored in order to rule out various
scenarios, including several where absorbers are largely galactic as well as 
nongalactic.  Absorbers are likely to arise from some combination of 
scenarios described below, such as disks, halos, and gas between galaxies.
Thus the results of the tests described here are likely to vary with 
limiting equivalent width or neutral column density.  For example, the
weakest absorbers, seen down to $N_{HI}\sim 10^{12}$ cm$^{-2}$ are less likely
to be associated with particular galaxies compared to stronger lines
such as those seen by the HST Key Absorption Line Project 
(Bahcall et al.~1996)
which typically have $N_{HI}>10^{14.3}$ cm$^{-2}$.

\begin{figure}[tb]
\plotfiddle{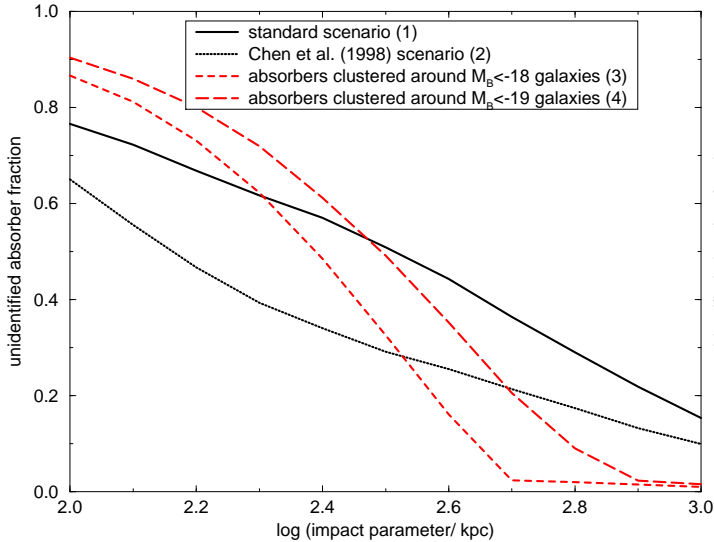}{2.3in}{270}{45}{45}{-185}{255}
\vskip -0.1in
\figcaption[fig5.eps]{The fraction of absorbers for which no galaxy is 
identified (UAF),
using the selection criteria as in Fig.~2, within some impact parameter
is plotted versus impact parameter for galactic absorbers as in the standard
scenario (1,solid), for galaxy halo absorbers obeying equation
24 of Chen et al.~1998 (2,dotted),
for weak absorbers which are clustered (within 750 kpc) around HSB,
$M_B<-19$ galaxies (3, dashed), and for weak absorbers clustered (within 500
kpc) around HSB $M_B<-18$ absorbers (4, long-dashed).
\label{fig5}}
\end{figure}

\begin{figure}[tb]
\plotfiddle{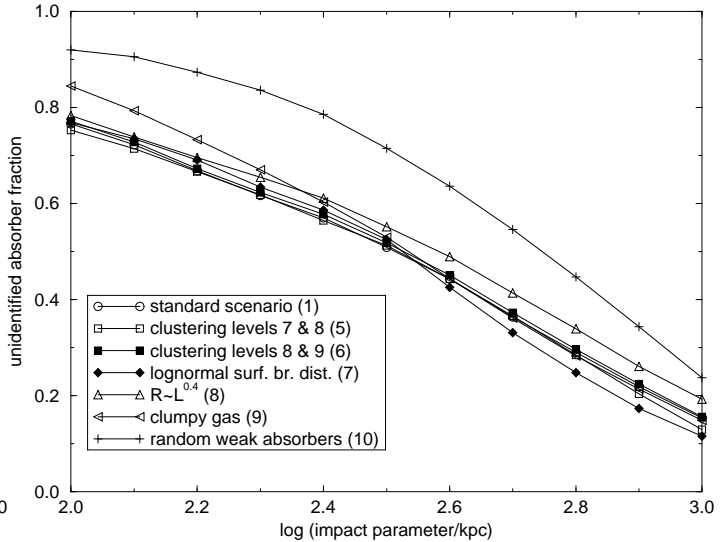}{2.3in}{270}{45}{45}{-185}{255}
\vskip -0.1in
\figcaption[fig6.eps]{The unidentified absorber fraction (UAF) is shown
versus impact parameter again for the standard scenario as in 
Fig.~5.  Here the standard scenario (circles) is compared with other 
galactic absorber scenarios including those with variations in the 
number of clustering levels (5, squares and 6, filled squares),
a lognormal galaxy central surface brightness distribution (7, diamonds), a less
steep Holmberg relation (8, triangles), clumpy gas (9, sideways triangles).
The UAF is also shown for weak absorbers that are distributed randomly 
relative to galaxies (plus symbols), in which case 
the UAF is consistently larger than for galactic absorbers.  The UAF slope
is determined largely by the slope of the autocorrelation function for observed
galaxies, so that little variation is seen for galactic or random absorbers.
Variations in the UAF amplitude will occur for various reasons, including 
uncertainties in the absorption cross sections and number density of galaxies.
\label{fig6}}
\end{figure}

\begin{figure}[tb]
\plotfiddle{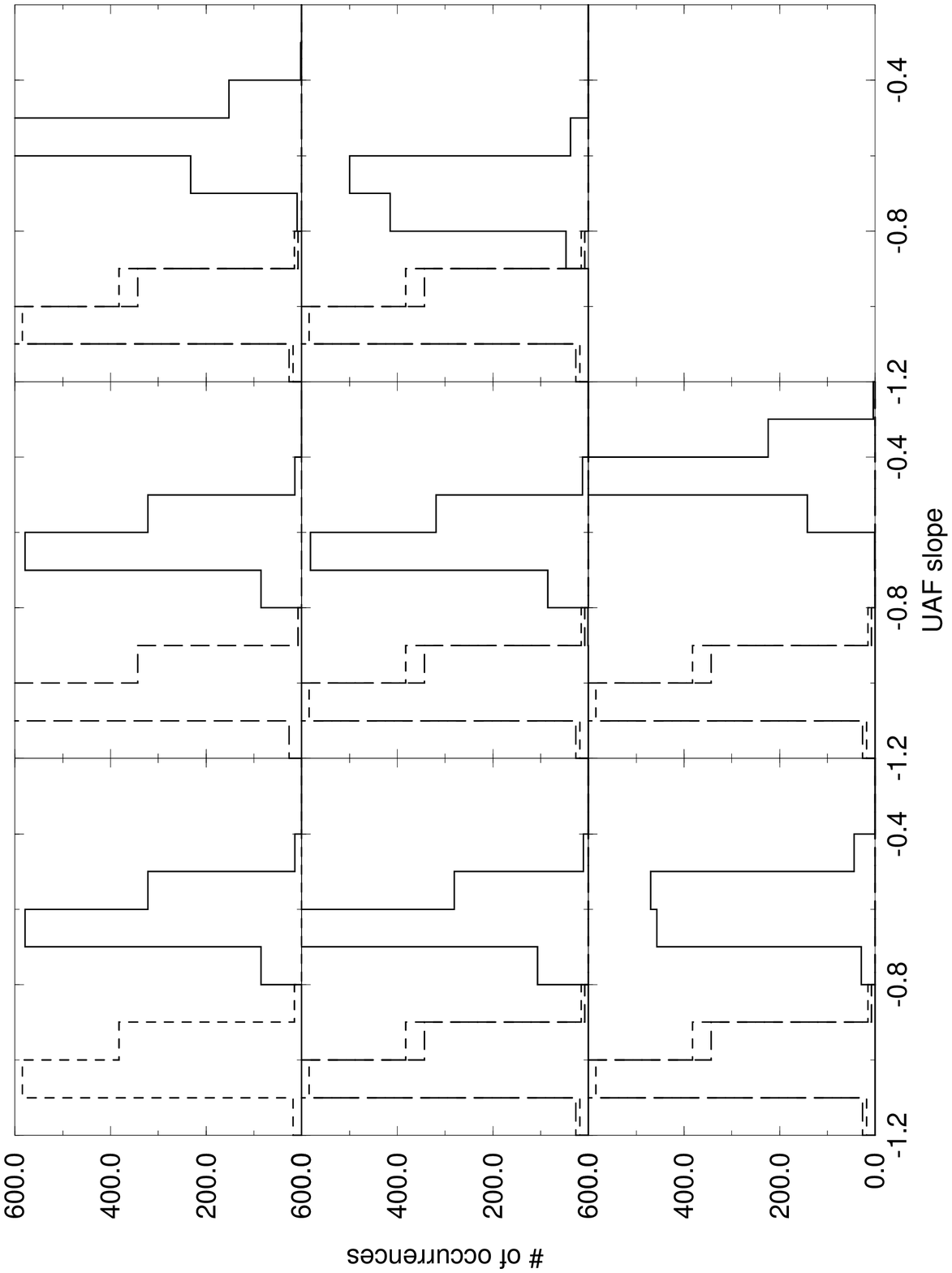}{2.3in}{270}{36}{36}{-155}{245}
\vskip -0.7in
\figcaption[fig7a.eps]{The UAF curves for 1000 samples of 100 galaxies, for
each of the scenarios shown in Figs.~5 and 6, were fitted to 
lines, and the binned slopes are shown in Fig.~7a.  In Fig.~7b, the
UAF slopes are shown for the same scenarios, now `observed' for $\triangle
V<750$ km/s, $M_B<-19$, and $\mu_B(0)<22$ mag arcsec$^{-2}$. The scenarios
shown, in both Figs.~7a and 7b, are from left to right, 1 and 3; 1 and 4;
2, 3, and 4; 5, 3, and 4; 6, 3, and 4; 7, 3, and 4; 8, 3, and 4; and 9. 3. and
4.  All galactic scenarios are shown as solid lines, and nongalactic 
scenarios are shown as dashed (3) and long-dashed (4).
For either type of selection criteria, an observed
sample of 100 galaxies is likely to allow for a conclusive test between galactic
and nongalactic absorber scenarios using the UAF slope.
\label{fig7}}
\end{figure}

\begin{figure}[tb]
\plotfiddle{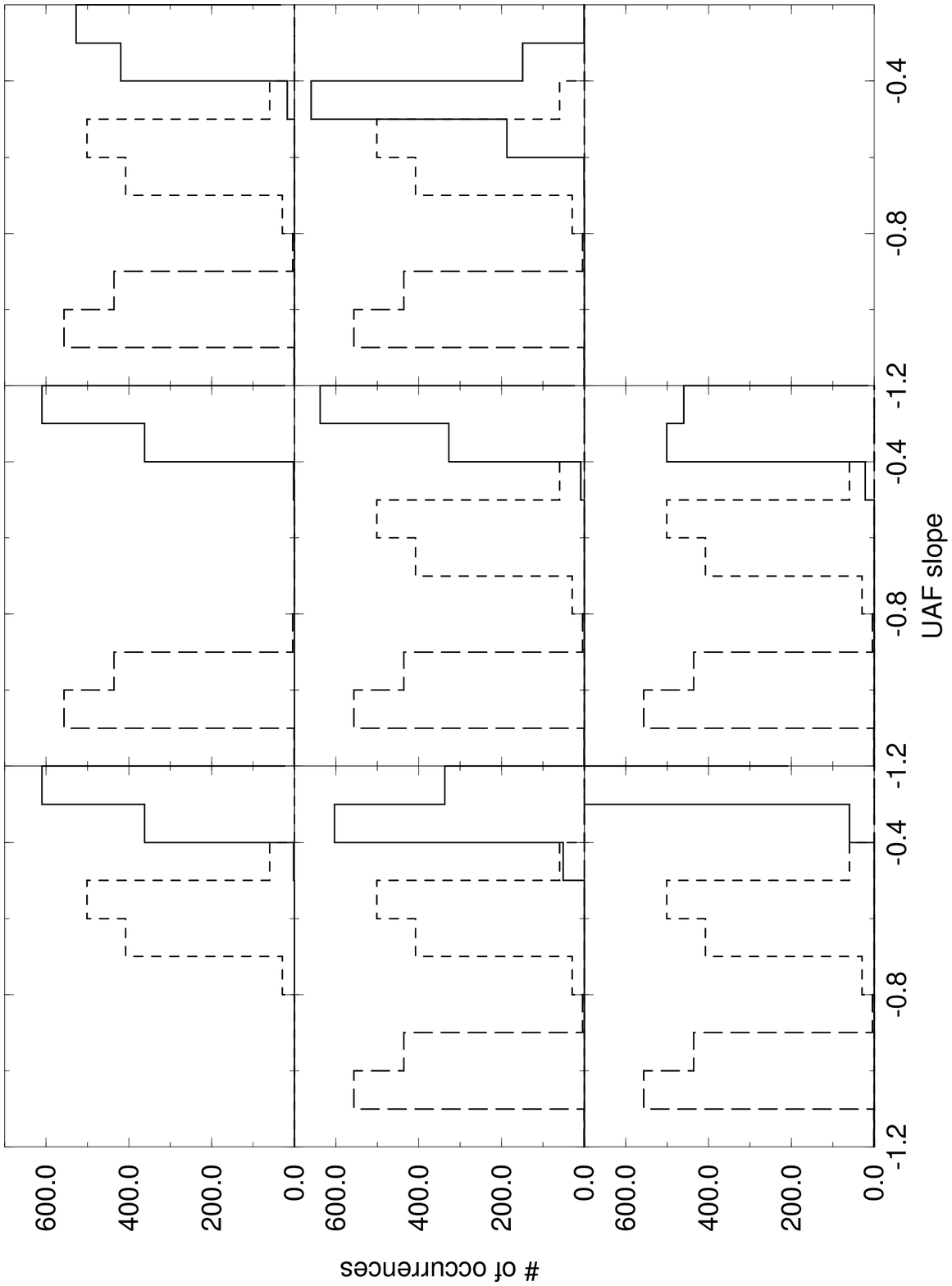}{2.3in}{270}{36}{36}{-155}{245}
\end{figure}

While it is never possible to be certain that a particular galaxy (if
any) gives rise to a particular absorption line, the fraction of absorbers
for which a galaxy is observed and the fraction of observed galaxies 
for which absorption is seen to arise will vary for galactic versus
nongalactic absorption scenarios.  The plots here are made for impact 
parameters $\ge 100$ kpc because weak absorbers are generally seen with
the nearest galaxy at a moderate to large impact parameter, while it
would be more difficult to simulate (or observe) sufficient numbers of 
galaxies for these tests to be meaningful at smaller impact parameters.

The fraction of absorbers for which no galaxy is identified (or unidentified
absorber fraction, UAF) within some impact parameter is plotted versus impact
parameter in Fig.~\ref{fig5}, 
using the same selection criteria as in the previous
section.  For nongalactic absorbers, the curves fall off much
more steeply compared to galactic absorbers.  If most absorbers are not 
associated with particular galaxies, then it is unlikely that a galaxy 
will be seen very close to the quasar line of sight.  Since the nongalactic
absorbers are assumed to be clustered around luminous HSB galaxies, it
becomes quite likely that a galaxy will be detected within several hundred
kpc of a line of sight.  The flattening at large impact parameters
is related to the clustering behavior of the gas around the galaxies.
The absorbers 
simulated using (scenario 2) the model of Chen et al.~(1998) have a 
consistently lower
UAF compared to the standard scenario.  This 
difference occurs because a larger fraction of absorbers arise in galaxies
that can be detected in scenario 2, as it was assumed
that absorption cross sections are unrelated to surface brightness.  In any
galactic absorber scenario there is some chance that an actual absorbing
galaxy is not identified, but then some other galaxy may be found at a 
larger impact parameter.  Thus the slope for the UAF is determined largely
by the clustering properties of galaxies for galactic absorber scenarios,
whereas the slope of the UAF is determined more by the clustering behavior
of gas around galaxies for the nongalactic absorber scenarios.  Therefore,
varying the surface brightness distribution or the Holmberg relation, or
making the gas clumpy, has little effect upon the results of the UAF test,
as is seen in Fig.~\ref{fig6}, except when substantial changes are made
in the absorption cross sections of galaxies.  For example, in the lognormal
surface brightness distribution scenario where $\mu_B^*=21.65$ mag 
arcsec$^{-2}$, the majority of galaxies are compact and high in surface
brightness, so that the number density of galaxies must be increased by a
factor of 3.5 in order to explain absorber counts.

The scenario where absorbers are distributed randomly relative to galaxies
would produce a UAF slope that would be indistinguishable from those in 
galactic absorber scenarios.  Clustering absorbers
with a wider range of column densities (by including absorbers $<10^{14.3}$
cm$^{-2}$) out to larger distances around luminous HSB galaxies would produce
a similar effect.  These scenarios are unlikely, however.  Given that most 
absorbers
have been seen to trace the large scale galaxy distribution, it is reasonable
to expect, on average, some falloff in column density with distance from the
center of a cluster.  Some evidence is seen that the weakest absorbers are
more likely to arise in void regions (Dav\'e et al.~1999; Shull et al.~1996;
Grogin \& Geller 1998),
and such a falloff would produce the equivalent width-impact parameter
anticorrelation seen out to large impact parameters by Tripp et al.~(1998).
Thus the weakest absorbers are likely to arise at large impact parameters from 
luminous galaxies, and for nongalactic absorbers it is reasonable to 
expect a UAF slope which is steeper than that for galactic or random absorbers.

It should 
be possible to distinguish between the galactic and nongalactic absorber
scenarios by observing a sample of 100 galaxies with well-defined selection
criteria, as shown in Figs.~\ref{fig7}.
For each scenario, the UAF slope is found for 1000 samples of 100 galaxies.
Little overlap in UAF slope is found for galactic versus nongalactic scenarios,
when sufficiently faint galaxies are detected.  Even when more luminous, 
higher surface brightness galaxies are detected, very little overlap occurs
except in the lognormal surface brightness distribution scenario (7).
The only other way to vary the UAF slope for galactic absorber scenarios would
be to vary the clustering behavior of galaxies.  For galactic absorber
scenarios the UAF slope is determined largely by the slope of the 
autocorrelation function for galaxies that are observed.  For the 
standard scenario the slope of the HSB autocorrelation is found to be
1.9 which is close to the value of 1.77 found for the CfA survey by
Davis \& Peebles (1983).  Varying the number of clustering levels
may cause some variation in the UAF normalization (although little change is
seen for the variations (scenarios 5 and 6)
 shown in Fig.~\ref{fig6}.), but the slope of the
autocorrelation function should remain the same for an adequately large
region of space (Soniera \& Peebles 1978).

The UAF test may be less conclusive in distinguishing between 
different galactic absorber scenarios.  The flattening at large impact
parameters may make it easier to constrain nongalactic absorber scenarios,
but uncertainties in the UAF normalization may occur in either case.
Obtaining a sample of $\sim 100$ galaxies may require looking out to a
high enough redshift that substantial uncertainties will exist in the 
normalization of the galaxy luminosity function or in the evolution of 
the relationship between galaxy luminosity and absorption cross section.
Some uncertainty in the UAF normalization will also occur as a result of 
uncertainties or evolution in the clustering behavior.

\begin{figure}[tb]
\plotfiddle{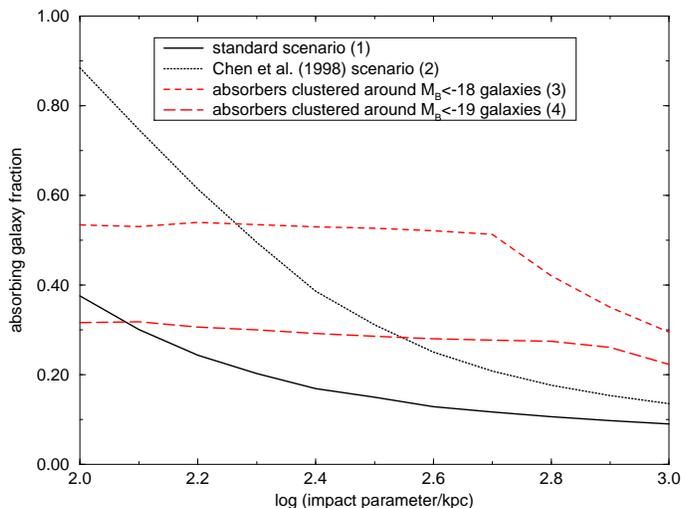}{2.3in}{270}{42}{42}{-175}{235}
\vskip -0.1in
\figcaption[fig8.eps]{The AGF, or fraction of galaxies with $M_B<-16$ and $\mu_
B(0)>23$ mag arcsec$^{-2}$ for which an absorption line ($>10^{14.3}$ cm$^{-2}$)
is seen within 400 km/s within some impact parameter, is shown versus 
impact parameter for the standard
scenario (1, solid), the Chen et al.~(1998) absorber scenario (2, dotted), and 
nongalactic absorber scenarios  (3, dashed and 4, long-dashed).  A larger
fraction of easily visible galaxies appear to give rise to absorption at 
large impact parameters when
the absorption arises in gas clustered around easily visible galaxies.
Absorption is seen to arise in galaxies least often when the absorption 
actually arises in unseen galaxies and in extended disks as in the standard
scenario.
\label{fig8}}
\end{figure}

Plots showing the absorbing galaxy fraction (AGF)
within some impact parameter may allow
for additional constraints in distinguishing between different absorber
scenarios.  
Again, note that even for galactic absorber scenarios, an 
observer is often seeing absorption that does not arise in the particular
galaxy that is detected.  Yet differences should occur in AGF
plots depending upon whether the absorbing gas is associated with particular
galaxies or with the large scale galaxy distribution.  Defining and simulating
the selection criteria used will be necessary in comparing observations with
AGF plots as well as UAF plots.  For example, the number of 
galaxies seen will vary widely with limiting luminosity, which will affect
the fraction of galaxies seen that appear to give rise to absorption.

AGF curves for galactic and nongalactic absorber scenarios (1 through 4)
are shown
in Fig.~\ref{fig8}, where the galaxies are `observed' using the same selection 
criteria as above, but allowing multiple galaxies to contribute to an 
absorption line.  If absorbers arise in gas that is clustered around luminous
HSB galaxies,  then a large fraction of the galaxies that we
see will appear to give rise to absorption even out to large impact parameters.
This is because we are likely 
to see the galaxies around which the gas is clustered.  Thus the AGF curves
will tend to fall off more quickly for galactic absorber scenarios, where 
dwarf and/or LSB galaxies give rise to some absorption.  The AGF tends to
be smallest for the standard scenario, as the galactic absorbers arise in
disks as compared to the spherical geometry used in the Chen et al.~(1998)
scenario (2).

\section{Tests on Current Observations}

It may be possible to distinguish between galactic and nongalactic absorbers
using a sample of 100 galaxies with well-defined selection criteria, as
was shown in Section 4.  A large enough data set to make a strong conclusion
is not currently available, but the complications that may arise when
the UAF and AGF tests are implemented are discussed below.

The largest set of absorber/galaxy observations currently available is from 
the study by
Chen et al.~(1998).  In this study galaxies were found within $\sim 200$ kpc of
quasar lines of sight, assuming that absorbers arise in particular galaxies
with sizes on that order.  These observations are less useful in testing
for the existence of nongalactic absorbers which might be seen when galaxies
are at larger impact parameters from a line of sight.  The UAF and AGF curves
are generally quite flat at such impact parameters for any scenario, so it
would be difficult to distinguish between slopes at these impact parameters.

Numerous galaxies, including some that are quite low in luminosity, have
been detected near quasar lines of sight by Bowen et al.~(1996) and Bowen
et al.~(1998).  These galaxies are located at a wider range of impact parameters
from the lines of sight, and they are at very low redshifts where evolution
in the galaxies and in the nature of absorbers would be negligible.
Unfortunately few absorbers in the column density range of interest here
have been detected.  Other studies that attempt to identify very weak
absorbers such as Tripp et al.~(1998) also include few absorbers that are
as strong as those simulated here.
\begin{figure}[tb]
\plotfiddle{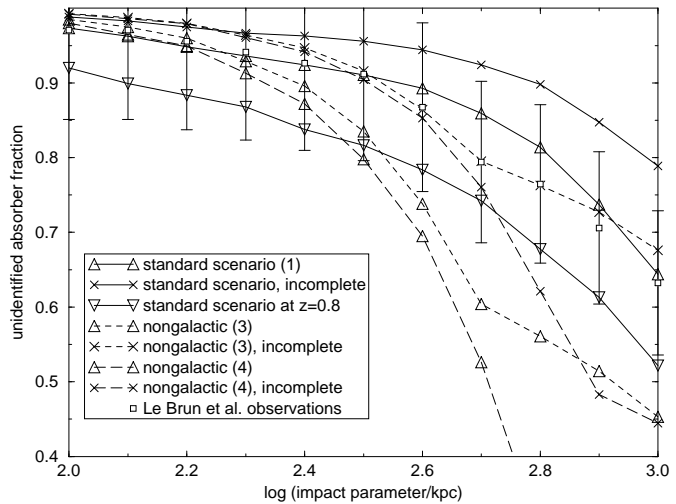}{2.3in}{270}{42}{42}{-175}{235}
\vskip -0.1in
\figcaption[fig9.eps]{UAF curves for `observed' simulations, where $M_B<-19$,
$\mu_B(0)<22$ mag arcsec$^{-2}$, and $\triangle V<750$ km/s, are shown for
the standard scenario (1, solid line), and nongalactic absorber scenarios 
(8, dashed and 9, long-dashed).
For each scenario incomplete observations are simulated.  UAF 
curves are shown where all the galaxies satisfying the selection criteria
are observed (triangles) and where half of such galaxies are observed
(X symbols). The UAF curve is also shown for the standard scenario where all of such 
galaxies are observed at $z=0.8$ assuming $q_0=0.1$ (triangles down).   These
UAF curves are compared with observations (filled circles) from Le Brun et 
al.~(1996) and Le Brun \& Bergeron (1998).  These observations may be 
consistent with most of the scenarios discussed so far, although the most
extreme nongalactic scenarios, where absorbers arise in gas clustered around
only luminous HSB galaxies such as those seen in these observations, appear 
to be unlikely.
\label{fig9}}
\end{figure}
\begin{figure}[tb]
\plotfiddle{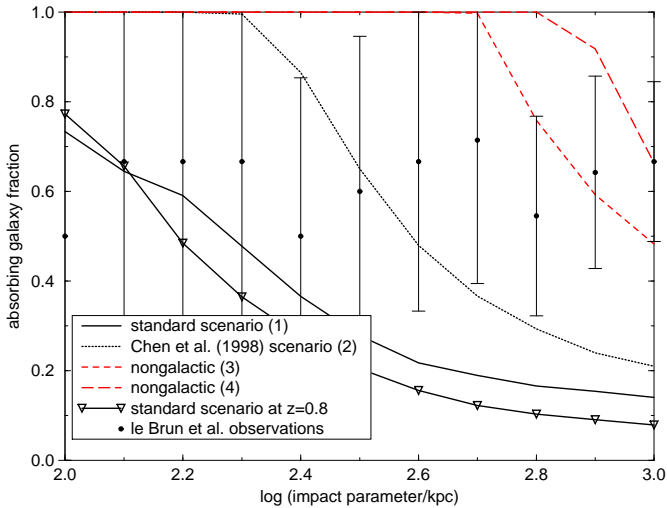}{2.3in}{270}{42}{42}{-175}{235}
\vskip -0.1in
\figcaption[fig10.eps]{AGF curves are shown, using the same selection criteria
as in Fig.~9, for the standard scenario (1, solid), the Chen et al.~(1998) 
scenario (2, dotted), and the two nongalactic absorber scenarios (8, dashed and
9, long-dashed).  The AGF curve for the standard scenario at $z=0.8$, where $q_0=0.1$,
is also shown.  These AGF curves are again compared with the observations from  
Le Brun et al.~(1996) and Le Brun \& Bergeron (1998).  The observations appear
to have little resemblance to any of the scenarios shown here, although the 
error bars are large, but reasons for a possible lack of resemblance are
discussed in the text.
\label{fig10}}
\end{figure}

Plots of the UAF and AGF for absorbers $>0.24$ \AA\ are shown in 
Figs.~\ref{fig9} and \ref{fig10} for the observations 
from Le Brun et al.~(1996) and Le Brun \& Bergeron (1998).  These samples
contain about 68 absorbers (although some have rest equivalent widths below
$0.24$ \AA) and about 28 galaxies with known redshifts that are in the range
in which Ly$\alpha$ absorption could be detected.  
While some of the galaxies in these samples are as
faint as $M_B\sim -19$, a limiting apparent magnitude is used for galaxies with
a wide range of redshifts up to $\sim 0.8$, and redshifts were not obtained for
all of the galaxies.  In the future, the simulated galaxy population can
be moved to varying redshifts within the observed redshift space in order to 
simulate galaxies with a wide range of properties that are detected using
a limiting apparent magnitude.  Cosmological surface brightness dimming and
K-corrections will need to be included in the simulation, and it will be
necessary to simulate absorbing galaxies in the wave band that is observed,
thus using appropriate luminosity and surface brightness distributions.  A
large enough data set is not available to merit such a careful simulation
at this time.

In Fig.~\ref{fig9} the observations are compared with simulated UAF curves for the
standard scenario and the nongalactic absorber scenarios (3 and 4), 
where all or half 
of the galaxies satisfying the selection criteria are observed for each 
scenario in order to show the effects of incompleteness.  The actual level of
incompleteness
is unknown, although redshifts were not obtained for some galaxies close to
lines of sight in Le Brun et al.~(1996).  Incompleteness generally raises the
UAF curves.  It is also possible that a data set could be less complete at
larger impact parameters, which would make the UAF curve less steep.

The UAF curves shown so far have been simulated at $z=0$, although some of
the galaxies observed by Le Brun et al.~(1996) were at redshifts as large
as $\sim 0.8$.  It is possible to adjust the simulation to conditions at $z=0.8$
by adjusting the number density of galaxies to account for the expected 
increase in absorber counts ($dN/dz$), assuming no evolution in the absorber
populations, using equation (4.3) in Weedman (1986).  Supposing that 
evolution in $dN/dz$ arises largely from changes in the ionizing background
radiation (as seen by Dav\'e et al.~1999), the ionizing background 
radiation is increased by a factor of $4.4$ as in M\"ucket, Petitjean, \& 
Riediger (1997).  Increasing the ionizing radiation reduces the absorption
cross sections of galaxies,  but the remaining absorbers are then located
closer to galaxies, so that the UAF is decreased as seen in the Figure.
However, only more luminous galaxies would tend to be seen at higher redshifts,
which would cause a somewhat opposite effect.  

Given the large uncertainties
in the currently available data, the observations from Le Brun et al.~(1996)
and Le Brun \& Bergeron (1998) may be consistent with most of the scenarios
discussed so far.  However, it appears unlikely that the observations are
consistent with the more extreme nongalactic absorber scenario (3) where absorbers
are clustered around HSB galaxies with $M_B<-19$.  Thus it is likely that 
at least some absorbers arise in gas around or associated with galaxies that
are lower in luminosity and/or surface brightness.

AGF plots are likely to be less sensitive to incompleteness in galaxy 
observations, but more sensitive to how uniformly gas is distributed around
galaxies.  The AGF is plotted for the Le Brun et al.~(1996) and Le Brun \& 
Bergeron (1998) observations in Fig.~\ref{fig10}.  Only a couple of galaxies in these
samples clearly do not have absorption lines at similar redshifts, which may
be indicating that the weakest absorbers arise outside of particular galaxies,
yet some stronger lines clearly arise close to galaxies as well.  But again,
many lines have rest equivalent widths $<0.24$ \AA, and the nature of the 
moderately weak lines of interest becomes less clear.  

While the error bars are quite large for the data points shown in 
Fig.~\ref{fig10},
the observations seem not to resemble any of the scenarios simulated here,
although there may be several explanations for this.  Gas may not be 
uniformly distributed around galaxies, and a population of nonabsorbing 
galaxies may exist (in addition to the absorbing galaxy population).  One
possibility is that the gas within or between galaxies is arranged in 
extended sheetlike structures, as often seen in cosmological simulations.
Positioning galaxies with a reasonable correlation function may be more
difficult in this case with the methods used here, although crude attempts
have produced AGF curves that would be indistinguishable from those given
by the extended disk absorbers in the standard scenario.  

A population of nonabsorbing galaxies would lower the AGF normalization
if the galaxies were distributed randomly relative to absorbers.  It is
less clear how the AGF would be changed if an observer was looking at 
nonabsorbing galaxies in clusters.  If any absorption was seen, the 
observer would likely conclude that a large number of galaxies could be
contributing to a small number of absorption lines, thus raising the AGF.
Yet nonabsorbing galaxies are more likely to exist in clusters, where gas
is more likely to be more highly ionized and stripped away from particular
galaxies.  One example of a plausible nonabsorbing galaxy population 
would be elliptical galaxies, which often do not give rise to absorption
(whereas spirals almost always do) in the Chen et al.~(1998) study.

Another interesting possibility is that the absorption properties of galaxies
seen at large impact parameters could be different from those seen at
smaller impact parameters.  Given that galaxies are detected out to some
angular separation from a quasar line of sight (3.5' for Le Brun et al.~1996),
the galaxies detected at larger impact parameters are likely to be at higher
redshifts and also higher in luminosity and surface brightness.  AGF curves
fall off more slowly for more luminous galaxies.  Furthermore, while little
change is seen in the AGF for higher redshifts (adjusted for expansion and
evolution in the ionizing radiation), an absorber at a given neutral column
density may correspond to a smaller overdensity at a higher redshift (Dav\'e
et al.~1999) even though the gas is more highly ionized.  Thus galaxies at
larger impact parameters may be less likely to give rise to absorption at 
some limiting column density.  It is also possible that we are seeing some
other type of evolution in the absorber population, where we see less highly
ionized absorbers at higher redshifts.
Limiting angular separations used in 
absorber-galaxy
observations will need to be taken into account when simulating larger
data sets in the future.

\section{Conclusions}

Galaxy selection not only complicates our efforts in finding the properties
of the low redshift galaxy population, but it also affects our ability to
establish the nature of low redshift quasar absorption lines.  It is generally
not possible to be certain that an absorption line arises in any particular 
galaxy, so that absorption arising in dwarf or LSB galaxies may be attributed
to luminous HSB galaxies, typically at larger impact parameters from a quasar
line of sight.  Thus several hundred possibly absorbing galaxies would need
to be observed in order to test directly for the nature of moderately weak
Ly$\alpha$ absorbers and the properties of galaxies that typically give rise
to absorption.  

Less direct tests may be more useful for determining the nature of absorbers
with a somewhat smaller set of observations.  Observations may be compared
with simulated plots of the unidentified absorber fraction (UAF) and the 
absorbing galaxy fraction (AGF) versus impact parameter.  The simulated plots
must take into account various observational selection criteria.  These tests
would be easiest to use for a sample of absorbers and galaxies in a narrow 
range of redshift so that evolutionary effects would be negligible, although
it is also possible to simulate observed redshift ranges.  At very low 
redshifts galaxies that are low in luminosity and surface brightness would be
easiest to detect, but a sufficient number of absorbers have not been found
yet at such low redshifts.

Currently some strong absorbers are known to arise close to galaxies, and
arguments have been made that other strong absorbers arise in LSB galaxies
(Jimenez, Bowen, \& Matteucci 1999; Phillipps, Disney, \& Davies 1993).  Yet
the weakest detected absorbers tend to arise far from luminous HSB galaxies.
The anticorrelation between equivalent width  and impact parameter seen by
Chen et al.~(1998) and others gives some indication that the strongest 
($> 10^{16}$ cm$^{-2}$) absorbers are often associated with particular
galaxies, while the continuation of the same anticorrelation seen by Tripp
et al.~(1998) suggests that the weakest absorbers arise in gas surrounding
groups and clusters.  The question thus remains, at what column density 
or equivalent width do absorbers typically change from galactic to nongalactic?
Some evidence is seen, from using the UAF test, that moderately weak
absorbers are not 
clustered around only the most luminous, high surface brightness galaxies.
While not enough observations are available to make strong constraints on
the nature of such absorbers at this time, the tests proposed
here will be useful for answering this question in the future.

The actual relationship between absorbers and galaxies may, of course, be 
more complicated than the scenarios that have been simulated here.  Absorbers
may arise from a combination of galactic and nongalactic sources (and from
multiple components of galaxies) at a given column density, and the gaseous
extent of galaxies is likely to vary with field versus cluster environment.
Variations in the amount of ionizing radiation with environment may also be
important, and some evidence is seen that stars within a galaxy contribute 
to the ionization of the outer parts of the galaxy (Bland-Hawthorn, Freeman,
\& Quinn 1997).  Thus
LSB galaxies, dwarf galaxies, and any gas located outside of rich clusters
may make a more important contribution to absorption as a result of being 
less highly ionized.  The effects of a possibly lower dust content in LSB
galaxies are unclear, however.  Variations in the ionizing background 
radiation will be considered more carefully in the future.

The same kinds of tests proposed here may be useful in constraining the 
nature of weak metal lines such as MgII (Churchill \& Le Brun 1988).
It is not known yet whether such metals are distributed far from the centers
of luminous HSB galaxies, or more frequently in other types of galaxies.  
Finding out
how metals are distributed around different types of galaxies and how well
gas from the outer parts of galaxies is mixed with gas from the inner regions
will have strong implications for our future understanding of galaxy evolution.

\acknowledgments  I am grateful to J.~Charlton, G.~Bothun, C.~Churchill, 
R.~Ciardullo,
R.~de Jong, S.~McGaugh, and B.~Savage for helpful discussions.

\clearpage

\end{document}